\documentstyle[aps,prb]{revtex}

\begin{document}

\input epsf.sty
\twocolumn[\hsize\textwidth\columnwidth\hsize\csname %
@twocolumnfalse\endcsname

\draft

\widetext

\title{Spin Fluctuations in the Underdoped High-$T_{\rm c}$ Cuprate 
La$_{1.93}$Sr$_{0.07}$CuO$_4$
}

\author{H.~{\sc Hiraka}, Y.~{\sc Endoh}}
\address{Institute for Materials Research, Tohoku University,
Sendai 980-8577}
\author{M.~{\sc Fujita}}
\address{The Institute for Chemical Research, Kyoto University,
Uji 611-0011}
\author{Y.~S.~{\sc Lee}}
\address{National Institute of Standards and Technology, 
NIST center for Neutron Research, Gaithersburg,
Maryland 20899, U.S.A.}
\author{J.~{\sc Kulda}, A.~{\sc Ivanov}}
\address{Institut Laue-Langevin, 156X, 38042 Grenoble Cedex 9,
France}
\author{R.~J.~{\sc Birgeneau}}
\address{
Department of Physics, University of Toronto,
Toronto M5S 1A1, Canada
}

\date{\today}
\maketitle

%\vspace{-0.1in}

%%%%%%% abstract (below) %%%%%%%%%%%%%
\begin{abstract}
We performed magnetic inelastic neutron-scattering experiments
  on La$_{1.93}$Sr$_{0.07}$CuO$_{4}$ over a wide range of $\omega$ and
  $T$ ; $2 \leq \omega \leq 44$~meV and $1.5 \leq T < 300$~K.  The
  dynamic susceptibility $\chi^{\prime\prime}({\bf q},\omega)$ of this
  underdoped high-$T_{\rm c}$ superconductor ($T_{\rm c}=17$~K) is
  characterized by broad, incommensurate peaks. 
Here, the
  incommensurate wavevector $\delta_{\omega}$ is approximately 0.07
  reciprocal lattice units at low $T$ and $\omega$.  
The
  superconducting phase does not possess an observable gap in the spin
  excitation spectrum down to at least 2~meV.  Scaling behavior is
  demonstrated for the $Q$-integrated energy spectrum
  $\chi^{\prime\prime}(\omega)_{T}$ with respect to 
($\omega/{\rm
    k}_{\rm B}T$).  This scaling establishes a connnection between the
  magnetic excitations of the compositions on either side of the
  insulator-superconductor boundary in the La$_{2-x}$Sr$_{x}$CuO$_{4}$
  phase diagram.  We note a possible cross-over from an incommensurate to
  a commensurate response 
 for $\omega > 20$~meV or $T > 300$~K.
\end{abstract}
%%%%%%% abstract (above) %%%%%%%%%%%%%

\pacs{PACS numbers: }

%%%%% prb format (below) %%%%%%%%%%%%%%
\phantom{.}
]
\narrowtext
%%%%%%%% prb (above) %%%%%%%%%%%%%%%%%%

%%%%%%%% main text (below) %%%%%%%%%%%%%%
\section{Introduction}

Magnetic inelastic neutron-scattering (INS) experiments have been
performed to investigate the relationship between high-$T_{\rm c}$
superconductivity,~\cite{Kastner-Rev,Furrer-Ed} and the strong
two-dimensional (2D) spin fluctuations of the CuO$_2$ plane.  It is
well known that inelastic magnetic peaks appear at incommensurate 
positions in momentum space for La$_{2-x}$Sr$_{x}$CuO$_{4}$ (LSCO)
superconductors.  Recently, Yamada {\it et al.} systematically studied
the low-energy magnetic excitations by using crucible-free
high-quality single crystals.~\cite{Yamada-PRB} They found a direct
proportionality between the hole concentration ($x$), dynamic
incommensurability ($\delta_{\omega}$), and $T_{\rm c}$ for the
underdoped superconductors.  There is also a clear gap in the
low-energy spin excitations for optimally doped
La$_{1.85}$Sr$_{0.15}$CuO$_{4}$ and slightly over-doped
La$_{1.82}$Sr$_{0.18}$CuO$_{4}$.~\cite{Yamada-PRL,CH-Lee} These latter
compounds do not possess static magnetic order.

Recent elastic neutron-scattering experiments revealed that 
incommensurate static
magnetic correlations exist in the CuO$_{2}$ planes for both the
spin-glass insulators~\cite{Wakimoto-Direct,Matsuda-IC-C,Fujita} and
the underdoped superconductors~\cite{Kimura,Matsushita}.  A remarkable
feature is that the positions of the incommensurate peaks 
change dramatically as a
function of hole concentration at the insulator-to-superconductor
boundary of $x_{\rm cr} \simeq 0.055$.~\cite{Fujita} That is, the
incommensurate wavevector is approximately
parallel to the Cu-O-Cu direction in the
superconductors, but is at 45$^\circ$ to this direction in the
insulators.  At the special doping level of $x=0.12$, 
the incommensurate peaks are
resolution-limited, indicating that the spins are correlated over
distances exceeding
200~\AA~\
within the CuO$_{2}$ plane, and the peaks appear below
$T_{\rm c}$.~\cite{Kimura,Suzuki}.  For the materials with static
order, the inelastic scattering peaks occur at the same positions as
those of the elastic peaks within error.

Neutron experiments with high transferred energies have focused
primarily on antiferromagnetic LCO~\cite{Hayden-LCO,S-Itoh} and
optimally doped LSCO~\cite{Yamada-JPSJ,Hayden-pure-optimal}.  
To-date, few detailed studies of the high-energy spin fluctuations
have been
reported for LSCO superconductors near the insulator-superconductor
boundary of the phase diagram.  The non-superconducting spin-glass
compositions have been carefully studied by Keimer {\it et
  al.}~\cite{Keimer} and Matsuda {\it et al.}~\cite{Matsuda-JPSJ} for
La$_{1.96}$Sr$_{0.04}$CuO$_{4}$ and La$_{1.98}$Sr$_{0.02}$CuO$_{4}$,
respectively.  They observed commensurate spin fluctuations for
$\omega > 3$ meV and demonstrated scaling behavior of
$\chi^{\prime\prime}(\omega)$ with respect to the variable
($\omega/{\rm k}_{\rm B}T$) over a wide range of $T$ and $\omega$.

We have performed a systematic study of the spin fluctuations in
La$_{1.93}$Sr$_{0.07}$CuO$_{4}$ to explore the behavior of
$\chi^{\prime\prime}({\bf q},\omega)$ in a system with hole
concentration slightly above $x_{\rm cr}$.  We find that the
magnetic-excitation spectrum is gapless.  Also, the $\omega$ and $T$
dependences indicate scaling behavior similar to that in
the lightly-doped
spin-glass compositions.  There is possibly an 
incommensurate-to-commensurate 
transition at high energy transfers or high temperatures.
This paper is organized as follows.  Section 2 describes the
experimental procedures and results.  A discussion of the data is
given in \S~3, and \S~4 contains the conclusion.
%
%=======FIGURE INSERTION==================================
\begin{figure}
\centerline{\epsfxsize=3in\epsfbox{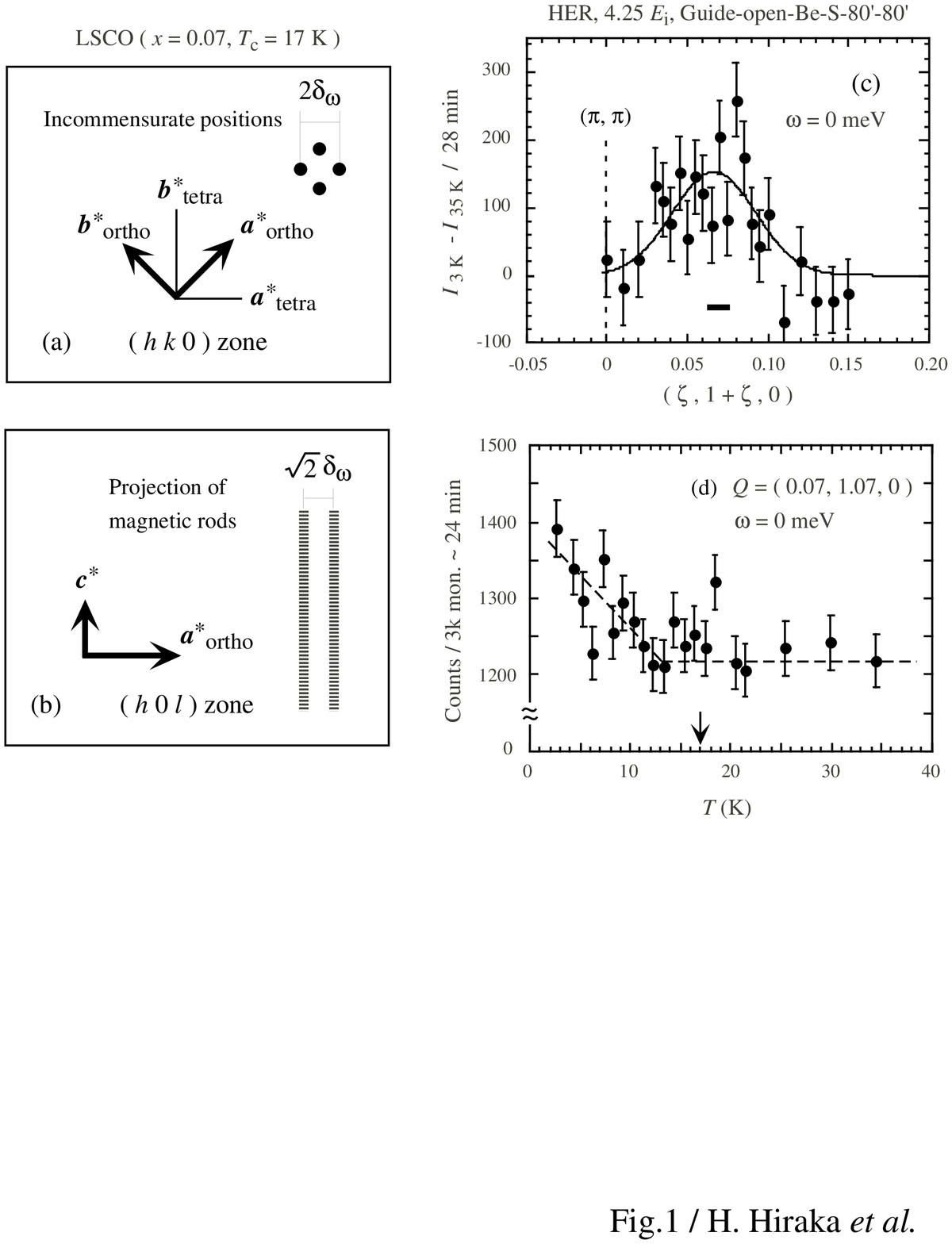}}
\caption{
(a) and (b): Scattering geometries in the ($h k 0$) and ($h 0 l$)
zones, respectively.
A magnetic four-rod structure is presented
(a) as the cross section giving four incommensurate points
on the tetragonal axes,
and (b) as the projection onto the 
({\boldmath $a$}$^{\ast}_{\rm ortho}$-{\boldmath $c$}$^{\ast}$)
plane with two parallel lines hatched along {\boldmath $c$}$^{\ast}$.
(c) and (d): Elastic scattering in the (a)-type geometry.
(c) An incommensurate peak in a difference plot 
between 3~K and 35~K.
The horizontal short bar represnts the instumental resolution.
(d) Thermal evolution of the peak intensity.
The arrow labels $T_{\rm c}$.
The broken line is a guide to the eye.
}
\label{fig:elastic}
\end{figure}
%=========================================================

\section{Experimental Procedure and Results}
The travelling-solvent-floating-zone method was utilized for the
single-crystal growth as described previously.~\cite{Hosoya-2,
  CH-Lee-FZ} The onset $T_{\rm c}$ of 17~K was determined from
dc-susceptibility measurements
and agrees with the linear relation between $T_{\rm
  c}$ and $\delta_{\omega}$.~\cite{Yamada-PRB} 
INS experiments were carried out on
the triple-axis spectrometers (TOPAN, IN8, IN20 and IN22) installed at
thermal-neutron beam ports of the JRR-3M in JAERI and of the high-flux
reactor in ILL.  
Either a pyrolytic-graphite (PG) monochromator or a copper monochromator
was used to monochromatize the indicent neutrons.  A PG analyzer was
used to select neutrons scattered from the sample.  
In most cases, a PG filter was
placed in front of the analyzer to remove higher-order neutrons from
the scattered beam.  
However, at IN20, we used a silicon analyzer utilizing
the (111) reflection without a filter.  The sample
(7~mm$\,\phi\times$ 30~mm) was set up in a He$^{4}$ closed-cycle
refrigerator or a pumped He$^{4}$ cryostat with He exchange gas.  A
twin structure exists in the low-temperature orthorhombic phase, in
which two domains are distributed with nearly equal populations.  The
orthorhombicity of ($b_{\rm ortho}/a_{\rm ortho}$) in {\it Bmab}
notation is 1.005 at 1.5~K.  Figures~\ref{fig:elastic}(a) 
and \ref{fig:elastic}(b) show the scattering
geometries in the ($h k 0$) and ($h 0 l$) zones, respectively.  We
conducted ${\bf q}$ scans along ${\bf q}_{\rm 2D}=[h,\pm~h,0]$ and
$[h,0,0]$ for each configuration, in order to pass through a pair of 
incommensurate peaks.

%\subsection{Cross-section}

The scattering function $S({\bf q},\omega)$ is directly connected to
$\chi^{\prime\prime}({\bf q},\omega)$ via the relation $S({\bf
  q},\omega)=(n+1)\chi^{\prime\prime}({\bf q},\omega)$ where
$n=\bigl({\rm exp}(\hbar\omega/{\rm k}_{\rm B}T)-1\bigr)^{-1}$.  
We
model the magnetic cross-section as four rods running along {\boldmath
  $c$}$^{\ast}$ in reciprocal space as illustrated in 
Figs.~\ref{fig:elastic}(a) and
\ref{fig:elastic}(b).~\cite{Thurston,Matsuda-SC} 
$\chi^{\prime\prime}({\bf q}_{\rm
  2D},\omega)$ is parameterized by four squared Lorentzians;
$\chi^{\prime\prime}({\bf q}_{\rm 2D},\omega)=
\chi^{\prime\prime}(\omega)\sum_{n=1}^{4} K_{\omega}^{2} /\bigl[ (
{\bf q}_{\rm 2D} - \mbox{\boldmath $\delta$}_{n}(\omega) )^{2}
+K_{\omega}^{2} \bigr]^2$.  Here, {\boldmath $\delta$}$_{n}(\omega)$
denotes the incommensurate wavevector and $K_{\omega}$ denotes the
$q$-width in the ({\boldmath $a$}$^{\ast}_{\rm ortho}$\,-\, {\boldmath
  $b$}$^{\ast}_{\rm ortho}$) plane.  $\chi^{\prime\prime}(\omega)_{T}$
is proportional to the ${\bf q}$ integration of
$\chi^{\prime\prime}({\bf q},\omega)$ over the entire Brillouin zone,
and the suffix $T$ emphasizes the temperature variation.  Instrumental
resolution effects~\cite{Cooper} were convolved with the above
cross-section.
%
%========FIGURE INSERTION=================================
\begin{figure}
\centerline{\epsfxsize=3in\epsfbox{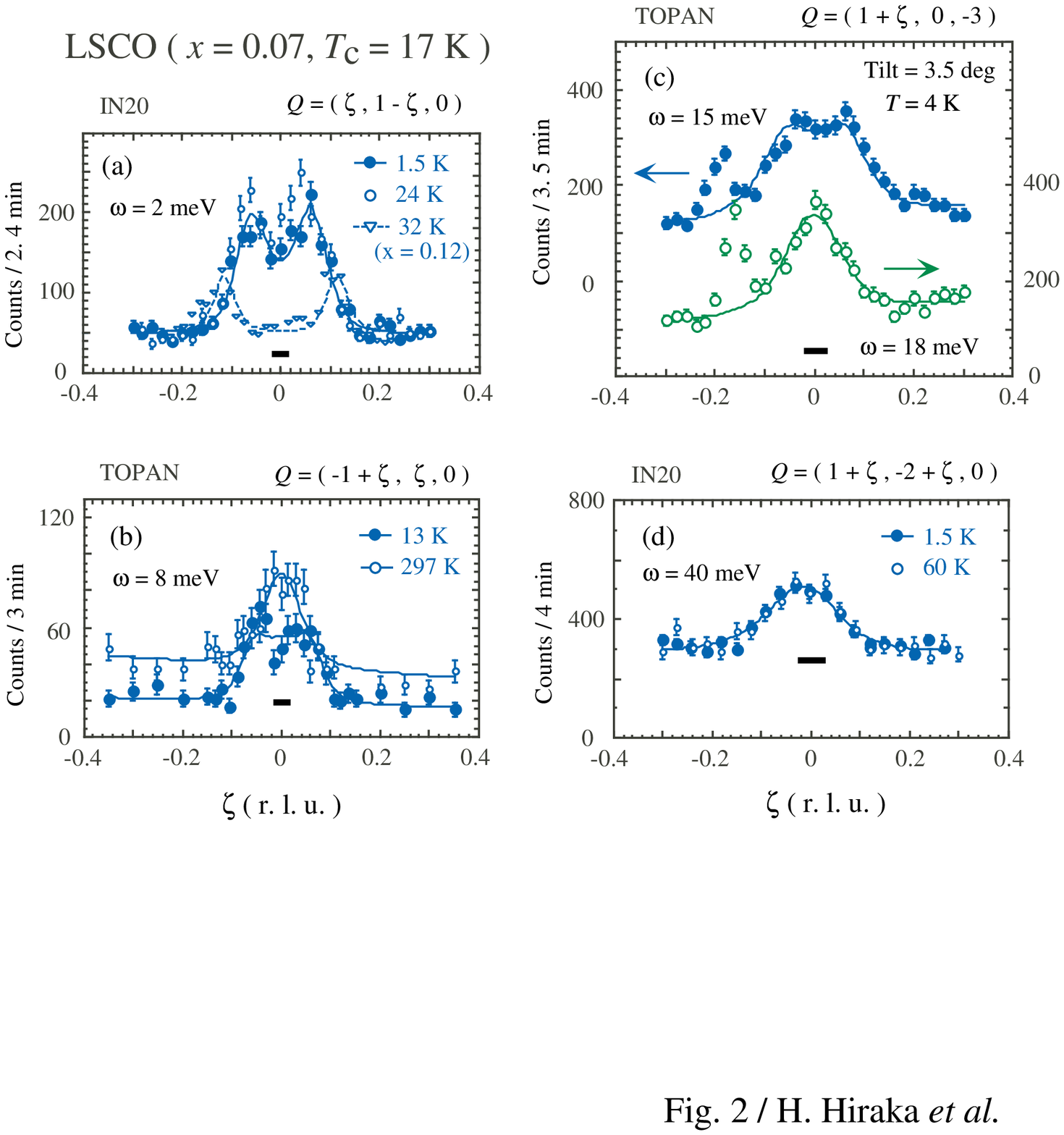}}
\caption{
(a)-(d): {\bf q}$_{\rm 2D}$ spectra at various $\omega$ and $T$.
Instrumental energies and collimations are as follows:
(a) 14.7~$E_{\rm f}$, Open-40'-40'-Open, 
(b) 14.7~$E_{\rm f}$, Open-30'-30'-Open,
(c) 13.5~$E_{\rm f}$, Open-60'-60'-Open
and (d) 18.7~$E_{\rm f}$, Open-40'-40'-Open.
The solid lines show the fitting results 
with resolution effects convoluted,
under a restriction in the peak width. 
The horizontal short bars represent the 
relevant instrumental $q$ resolutions.
As a reference, an incommensurate peak structure observed in
La$_{1.88}$Sr$_{0.12}$CuO$_{4}$ ($T_{\rm c}=31$~K) is also depicted
in (a) after normalization with 
the sample volume and monitor time.
The broken line is a guide to the eye.
}
\label{fig:q-spectra_E}
\end{figure}
%=========================================================

%\subsection{Elastic peak}
Before a detailed description of inelastic scattering measurements, we
briefly comment on the static correlations observed in this
sample.
Elastic neutron scattering was measured with the triple-axis spectrometer
(HER) placed in the cold-neutron beam line of JAERI.
At low $T$, magnetic elastic peaks
appear at incommensurate positions with $\delta_{0}\simeq 0.069 
a_{\rm tetra}^{\ast}$.
As shown in Fig.~\ref{fig:elastic}(c),
the static response is weak and broad in momentum
space ($\kappa_{0}^{\ 7\%}=0.037$~\AA$^{-1}$), compared to the
resolution-limitted peaks in
La$_{1.88}$Sr$_{0.12}$CuO$_{4}$ ($\kappa_{0}^{\ 12\%}\leq
0.005$~\AA$^{-1}$).~\cite{Kimura} The short-range static correlation
length ($1/\kappa_{0}^{\ 7\%}\leq 30$~\AA) and the small ordered
moment are consistent with the general trend found for superconducting
compositions with $x < 0.12$.~\cite{Kimura} The elastic scattering
merges into the background level above $\sim$ 12~K, slightly lower
than $T_{\rm c}$ as demonstrated in Fig.~\ref{fig:elastic}(d).
The detailed results and analyses of the elastic component
will be reported elsewhere.~\cite{Fujita}.
%
%\subsection{Raw spectra}

\begin{figure}
\centerline{\epsfxsize=3in\epsfbox{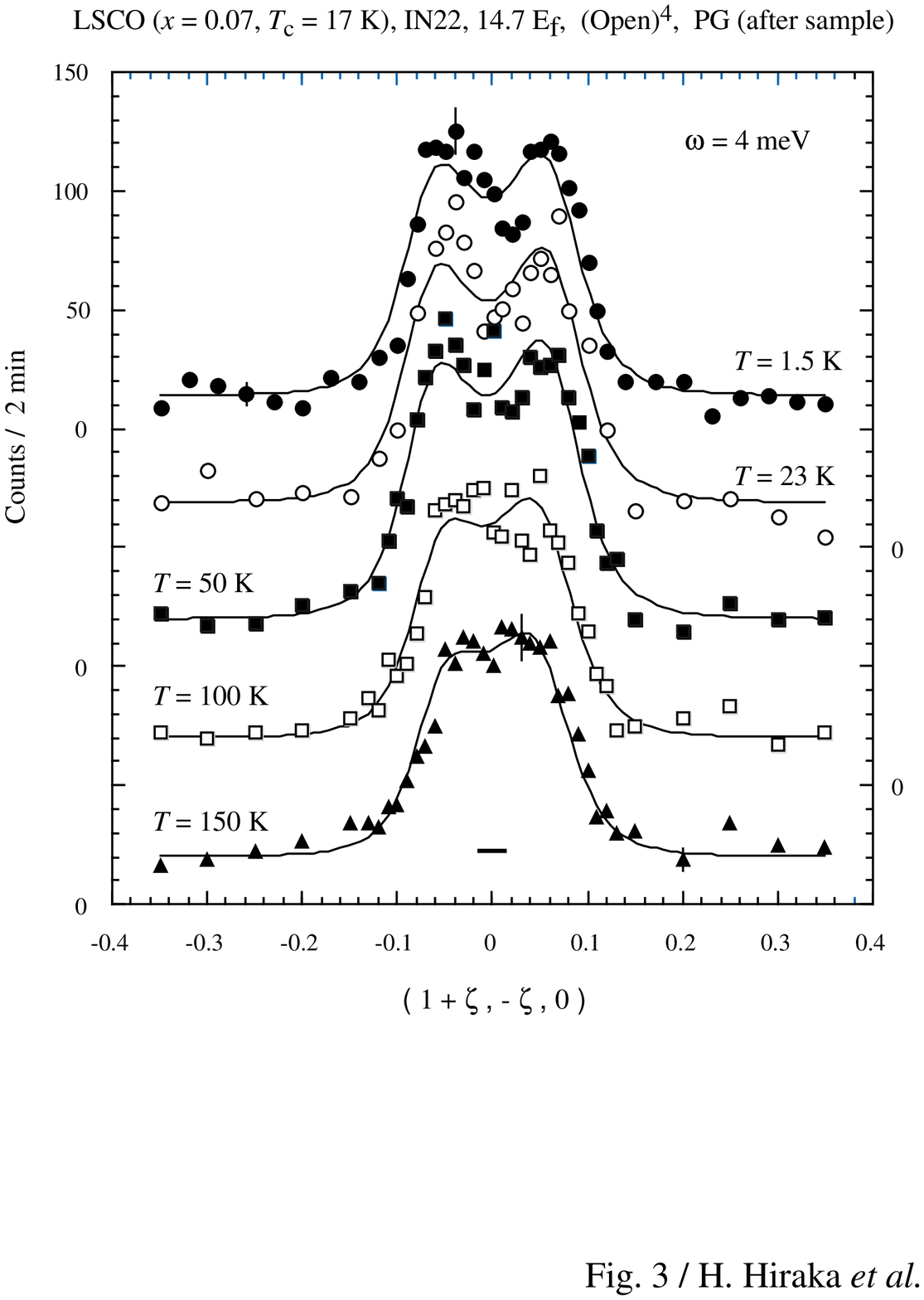}}
\caption{
Temperature variation of low-energy incommensurate peaks
with $\omega=4$~meV.
The solid lines show the results of resolution-convoluted
fits by using the same condition for the peak widths,
as the fits depicted in Figs.~\protect\ref{fig:q-spectra_E}(a)\,
-\,\protect\ref{fig:q-spectra_E}(d).
The horizontal short bar represents the instrumental $q$ resolution.
}
\label{fig:q-spectra_T}
\end{figure}

We display representative scans through the rods of magnetic
scattering for various energy transfers and temperatures in
Figs.~\ref{fig:q-spectra_E}(a)\,-\,\ref{fig:q-spectra_E}(d).  At low
$\omega$ and $T$, the inelastic scattering is incommensurate with
peaks at the same positions as those seen in the elastic channel.  
Pronounced
scattering survives even down to 2~meV at 1.5~K as shown 
in Fig.~\ref{fig:q-spectra_E}(a).
Therefore, in this compound any magnetic gap in the
superconducting state must be at energies much less than
2~meV.
At higher energy transfers and temperature, we
see evidence for a possible crossover from a double to a single peak
structure in the raw ${\bf q}_{\rm 2D}$ profiles shown 
in Figs.~\ref{fig:q-spectra_E}(b)
and \ref{fig:q-spectra_E}(c).  
Unfortunately, phonon scattering severely contaminates the
{\bf q}$_{\rm 2D}$ profiles in the medium-energy range of $12 < \omega
< 30$~meV, as seen at $\zeta \approx -0.2$~r.l.u. 
in Fig.~\ref{fig:q-spectra_E}(c).
Hence, the precise evolution of the line-shapes from the low energy to
the high energy regime is difficult to obtain.  
The temperature evolution of the magnetic scattering 
for $\omega = 4$~meV
is depicted in Fig.~\ref{fig:q-spectra_T}.  
The peak intensity and the incommensurate splitting
both appear to decrease gradually with increasing $T$.
%
%
%========FIGURE INSERTION=================================
%=========================================================

%
%\subsection{Analysis procedure}
%
For the initial data analysis, we allowed the parameters
$\delta_{\omega}, K_{\omega}, \chi^{\prime\prime}(\omega)$ and a
sloped background along ${\bf q}_{\rm 2D}$ to vary independently.
However, in the high-$\omega$ or high-$T$ region, the fits do not
converge uniquely; a large $K_{\omega}$ or a small $\delta_{\omega}$
can be invoked to describe the data.  In order to be systematic and
obtain physically consistent fits, we assumed a smooth increase in
$K_{\omega,T}$ with increasing $\omega$ and temperature:
$\bigl(K_{\omega,T}\bigr)^{2} = \bigl(K_{0,0}\bigr)^{2} + a_{0}^{-2}
\bigl[({\rm k}_{\rm B}T/E_{T} )^2 + (\omega/E_{\omega})^2\bigr]$ with
$K_{0,0}=0.065$~\AA$^{-1}$, $a_{0}=3.79$~\AA\ and
$E_{T}=E_{\omega}=85$~meV.  This form was used by Aeppli {\it et
  al.}~\cite{Aeppli-QCP} to explain the peak widths observed in
La$_{1.86}$Sr$_{0.14}$CuO$_{4}$.  The above expression for
La$_{1.93}$Sr$_{0.07}$CuO$_{4}$ describes the overall behavior of
$K_{\omega,T}$ fairly well. The fits to the {\bf q}$_{\rm 2D}$ spectra
are shown in Figs.~\ref{fig:q-spectra_E}(a)\,-\,
\ref{fig:q-spectra_E}(d) and \ref{fig:q-spectra_T}.  
$K_{0,0}$ corresponds to
$\kappa_{0,0}^{\ \ 7\%} = 0.042$~\AA$^{-1}$.  
The extrapolated value
is compatible with the value obtained for
$\kappa_{0}^{\ 7\%}$ from the elastic scattering, but it is
about twice as large as $\kappa_{0,0}^{\ \ 
  14\%}=0.022$~\AA$^{-1\,\,}$~\cite{Aeppli-QCP} and $\kappa_{\omega,T}^{\ 
  \ 4\%, 2.4\%}\sim0.02$~\AA$^{-1}$ 
found at low $\omega$ and low
$T$.~\cite{Keimer,Matsuda-IC-C} As a further comparison, 
Fig.~\ref{fig:q-spectra_E}(a)
also displays the sharp incommensurate peaks observed in
La$_{1.88}$Sr$_{0.12}$CuO$_{4}$ ($T_{\rm c}=31$~K), which are much
narrower than those we find in La$_{1.93}$Sr$_{0.07}$CuO$_{4}$
%
%\subsection{Integrated intensity}
%
%
%========FIGURE INSERTION=================================
\begin{figure}
\centerline{\epsfxsize=3in\epsfbox{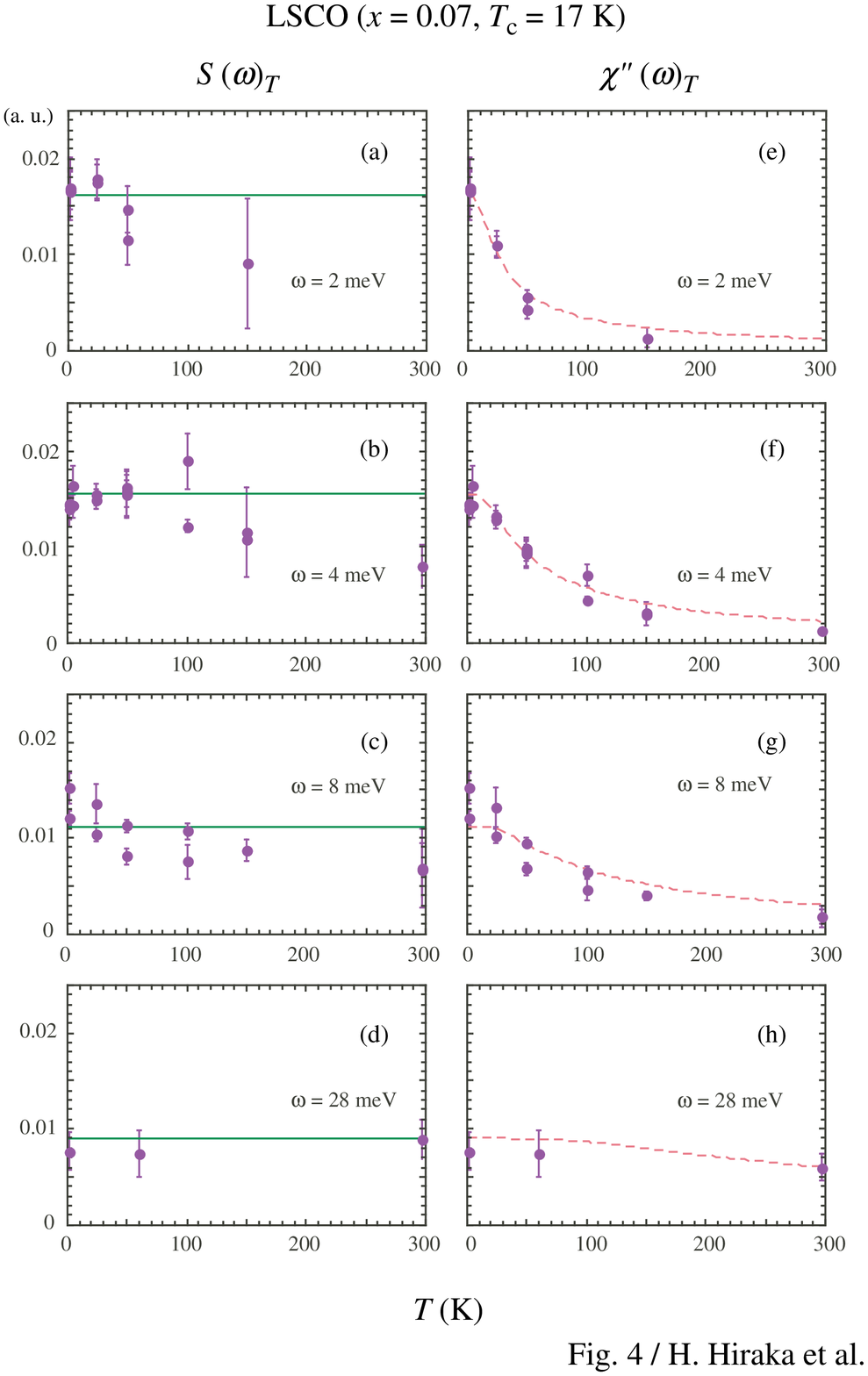}}
\caption{
Temperature dependence of $S(\omega)_{T}$ in (a)-(d) and 
of $\chi^{\prime\prime}(\omega)_{T}$ in (e)-(h).
The solid straight lines in (a)-(d) are drawn,
assuming that $S(\omega)_{T}$ is constant 
as a function of $T$,
while the broken curves in (e)-(h) are calculated
assuming
$\chi^{\prime\prime}(\omega)_{T}
=\chi^{\prime\prime}(\omega)_{1.5 {\rm K}}/
\bigl(n(\omega,T)+1\bigr)$.
}
\label{fig:S/chi_T}
\end{figure}
%=========================================================
%
%
%========FIGURE INSERTION=================================
\begin{figure}
\centerline{\epsfxsize=3in\epsfbox{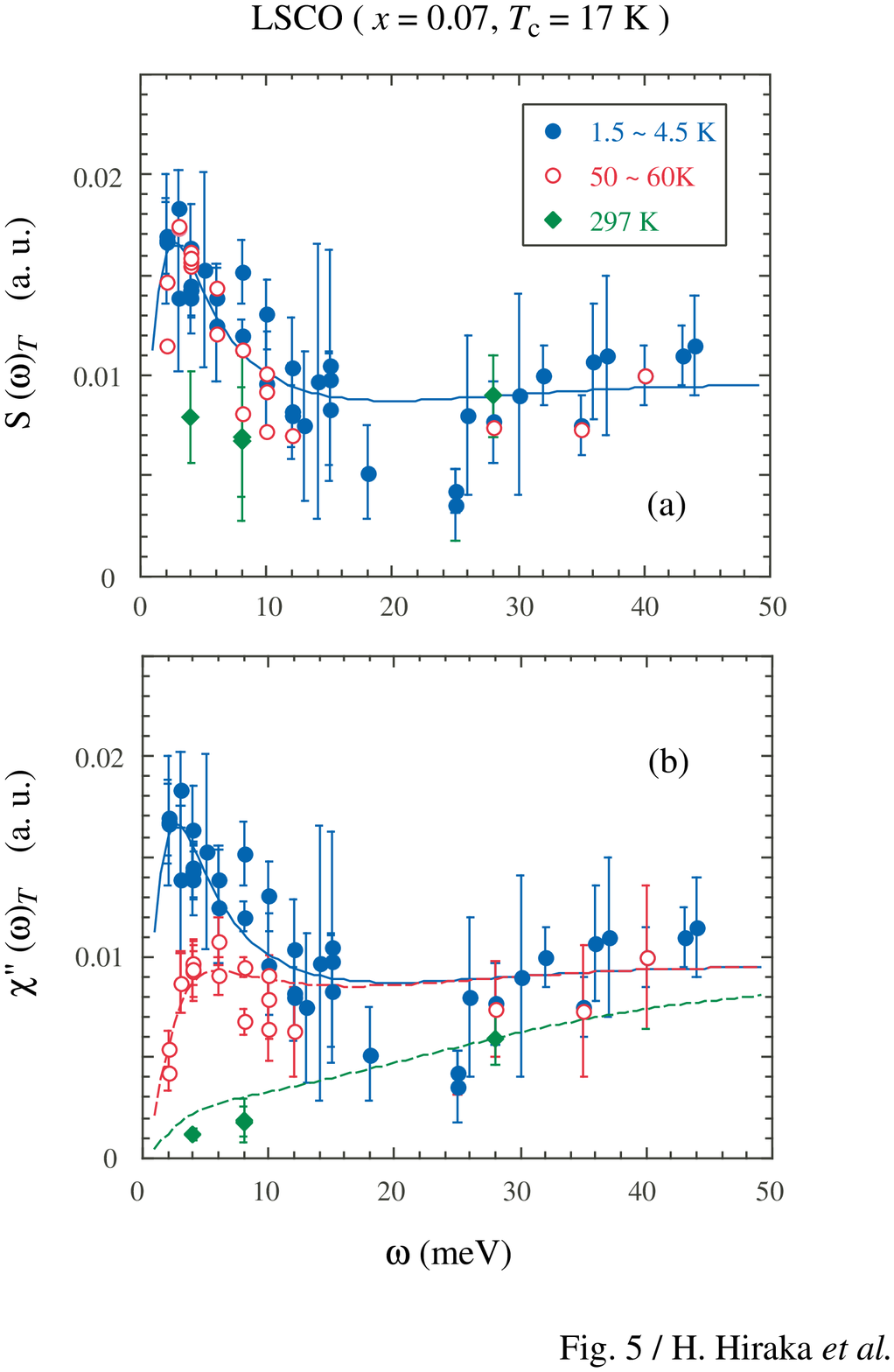}}
\caption{
$\omega$ dependence of (a) $S(\omega)_{T}$ and 
(b) $\chi^{\prime\prime}(\omega)_{T}$.
The solid line for $\chi^{\prime\prime}(\omega)_{1.5 {\rm K}}$
represents the addition of two putative damped Lorentzians;
$\omega \sum_{n={\rm L, H}} c_{n}{\sl\Gamma}_{n}/
(\omega^2+{\sl\Gamma}_{n}^2)$,
where
the characteristic energies ${\sl\Gamma}_{\rm L}$
and ${\sl\Gamma}_{\rm H}$ were arbitrarily chosen to be
2.5 and 60~meV, respectively, with the ratio of
coefficients $c_{\rm L}/c_{\rm H}$ of 2.
The broken lines in (b) show the thermal evoluition
calculated by
$\chi^{\prime\prime}(\omega)_{T}
=\chi^{\prime\prime}(\omega)_{1.5 {\rm K}}/
\bigl(n(\omega,T)+1\bigr)$.
}
\label{fig:S/chi_E}
\end{figure}
%=========================================================

\begin{figure}
\centerline{\epsfxsize=3in\epsfbox{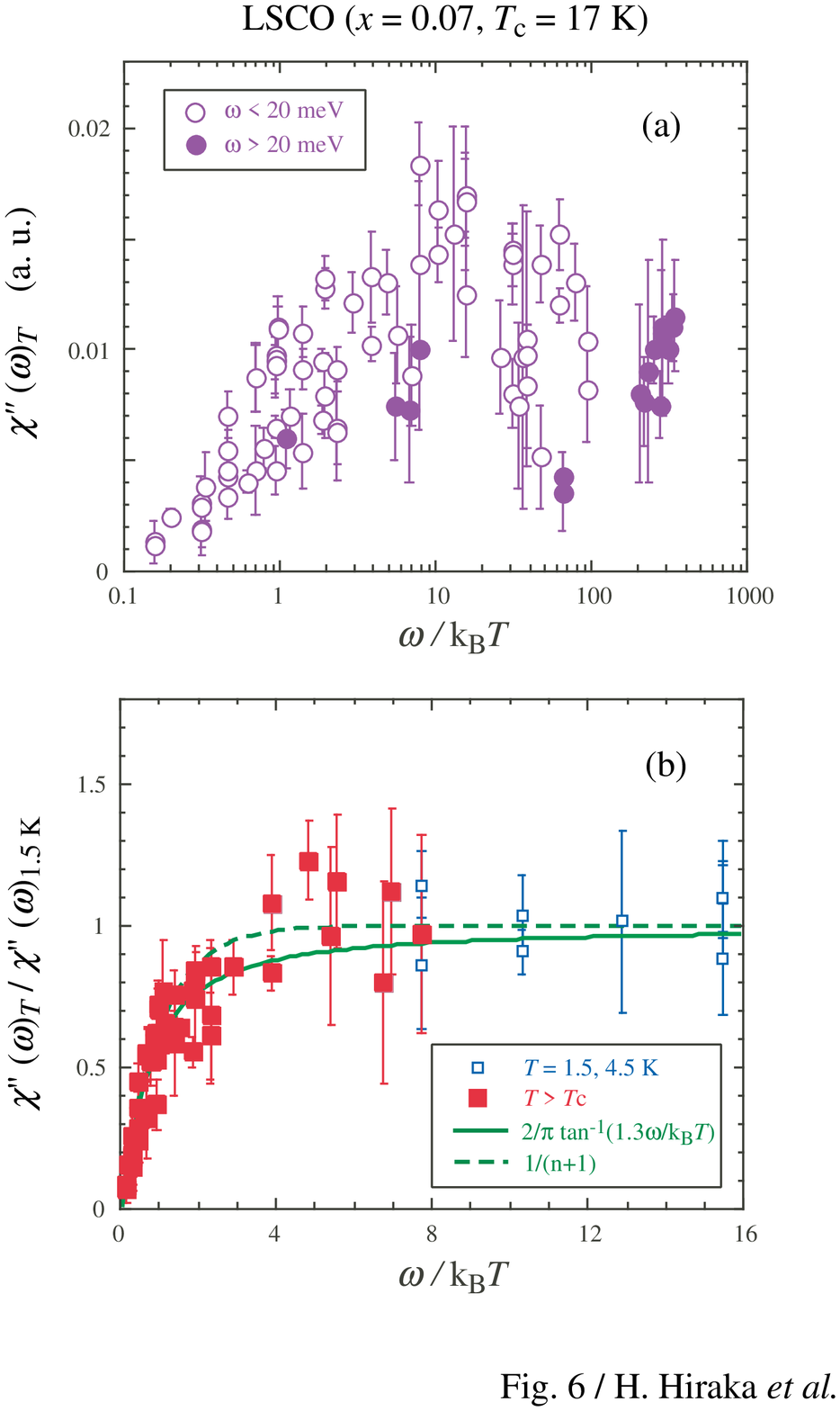}}
\caption{
(a) Direct replot of $\chi^{\prime\prime}(\omega)_{T}$
versus ($\omega/{\rm k}_{\rm B}T$).
Note that the horizontal axis is on a logarithm in scale.
(b) Conventional scaling plot for small ($\omega/{\rm k}_{\rm B}T$).
The solid and broken lines represent 
(2/$\pi$)tan$^{-1}\bigl(1.3(\omega/{\rm k}_{\rm B}T)\bigr)$
and $\bigl(n(\omega,T)+1\bigr)^{-1}$, respectively.
}
\label{fig:scaling}
\end{figure}

The $T$ dependences of $\chi^{\prime\prime}(\omega)_{T}$ and
$S(\omega)_{T}$ are depicted in Figs.~\ref{fig:S/chi_T} and 
\ref{fig:S/chi_E} for various values of $\omega$.
$S(\omega)_{T}$ is the {\bf q}$_{\rm 2D}$ integration of $S({\bf
  q},\omega)$, and depends weakly on $T$.  A remarkable feature of
$\chi^{\prime\prime}(\omega)_{T}$ is that low-energy fluctuations grow
significantly with decreasing $T$.  This is in contrast to the
complete gap seen in the excitation spectrum of the optimal
superconductor.  The data in Fig.~\ref{fig:S/chi_E} also indicate 
that the spectral
weight in the high-energy range ($>20$~meV) is nearly constant, or
tends to increase slightly with increasing $\omega$.  We plot
$\chi^{\prime\prime}(\omega)_{T}$ as a function of the scaled variable
($\omega/{\rm k}_{\rm B}T$) in Fig.~\ref{fig:scaling}(a).  The data
with $\omega < 20$~meV (open circles) roughly show a maximum at
($\omega/{\rm k}_{\rm B}T)\sim~10$.  This behavior is similar to that
in YBa$_2$Cu$_3$O$_{6+x}$ (YBCO) with $x=0.4$ located near the
antiferromagnetic-superconducting phase boundary.~\cite{Sato}
Higher-energy excitations (solid circles) are also shown.
Figure~\ref{fig:scaling}(b) shows $\chi^{\prime\prime}(\omega)_{T}$
divided by an $\omega$-dependent normalization factor measured at $T =
1.5$~K.~\cite{Keimer,Matsuda-JPSJ}
%
%
%========FIGURE INSERTION=================================
%=========================================================

\section{Discussion}
%\subsection{Energy spectrum,T-variation,Scaling}

Our main result is the determination of the magnetic-excitation
spectrum of La$_{1.93}$Sr$_{0.07}$CuO$_{4}$ for $2\leq \omega \leq
44$~meV for various temperatures.  At low temperatures, the upturn of
$\chi^{\prime\prime}(\omega)$ with decreasing energy transfer is
similar to the behavior observed in the non-superconducting,
spin-glass compositions.~\cite{Keimer,Matsuda-JPSJ} This indicates
that the low-energy spin fluctuations retain their 
same basic character
even as one crosses the insulator-superconductor boundary.  We note,
however, that the elastic scattering at the incommensurate positions
in La$_{1.93}$Sr$_{0.07}$CuO$_{4}$ is weaker than the strong
quasi-elastic scattering observed in La$_{1.96}$Sr$_{0.04}$CuO$_{4}$
and La$_{1.98}$Sr$_{0.02}$CuO$_{4}$.  $S(\omega)_{T}$ is largely
independent of temperature as seen in Fig.~\ref{fig:S/chi_T}.  
This indicates that
$\bigl(\chi^{\prime\prime}(\omega)_{T}
/\chi^{\prime\prime}(\omega)_{1.5~{\rm K}}\bigr)$ has a $T$-dependence
that roughly follows the inverse of the Bose thermal factor
$\bigl(n(\omega,T)+1\bigr)^{-1}$.  The scaling function tan$^{-1}(1.3
\omega/{\rm k}_{\rm B}T)$ also explains the ($\omega/{\rm k}_{\rm
  B}T$) dependence in Fig.~\ref{fig:scaling}(b) to within the errors.
The latter scaling function has been used to describe the data for
$\chi^{\prime\prime}(\omega)_{T}$ in La$_{1.96}$Sr$_{0.04}$CuO$_{4}$
and La$_{1.98}$Sr$_{0.02}$CuO$_{4}$ over a wide 
($\omega\geq 3$~meV,\,$T$) regime.~\cite{Keimer,Matsuda-JPSJ}
%
%%\subsection{Incommensurability}
%
%========FIGURE INSERTION=================================
\begin{figure}
\centerline{\epsfxsize=3in\epsfbox{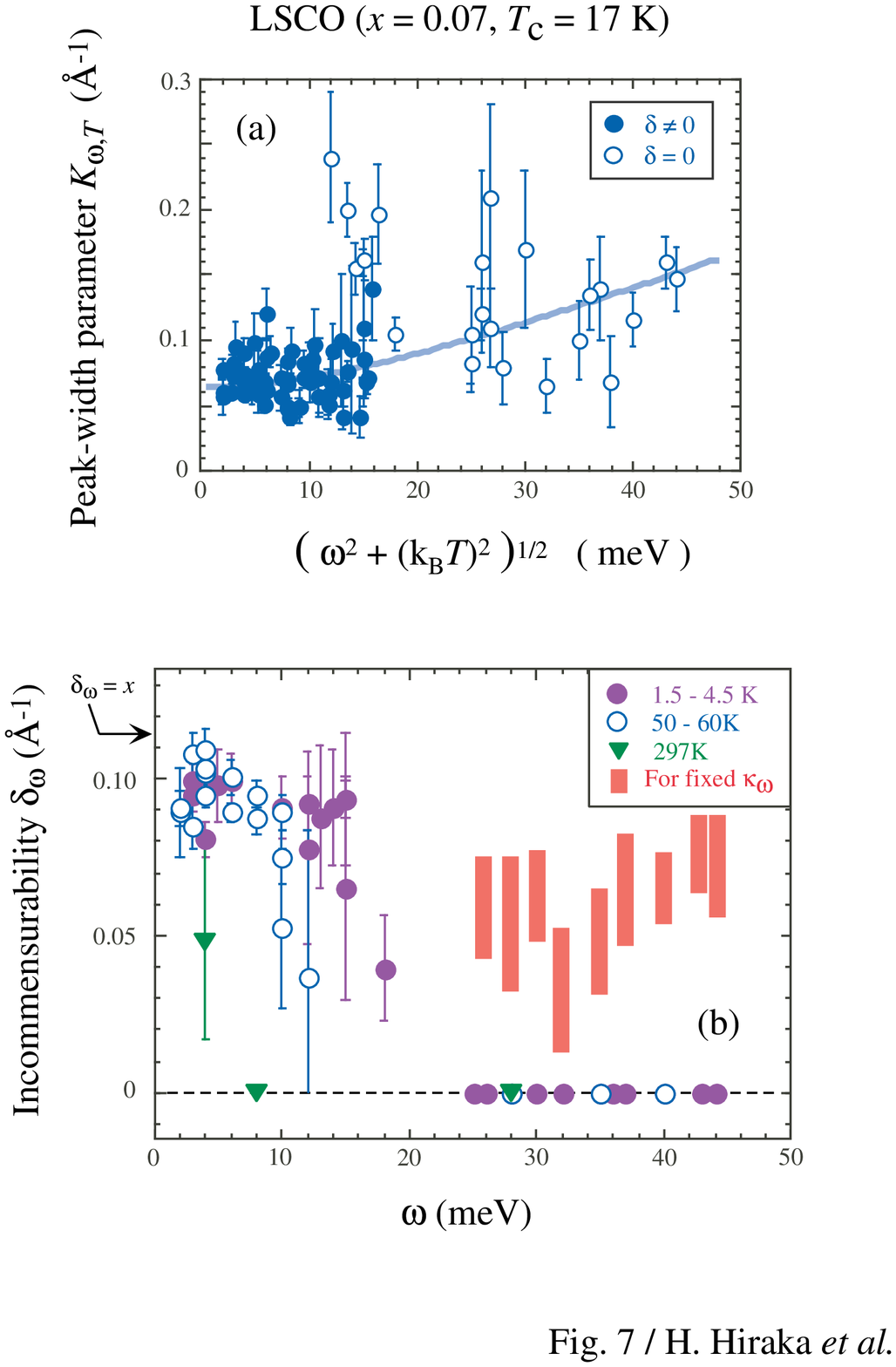}}
\caption{
(a) Peak-width parameter $K_{\omega,T}$ 
when no restriction is used in the {\bf q}-profile fits. 
A smooth increase of $K_{\omega,T}$ against $\omega$ and $T$
(a curved line)
is assumed in the main analysis. 
(b) Resolution-corrected incommensurability $\delta_{\omega}$.
Based on the above assumption for $K_{\omega,T}$,
a significant change of $\delta_{\omega}$ is deduced.
If the peak width is held constant 
($\kappa_{\omega}=0.04$~r.l.u. $=0.047~$\AA$^{-1}$,
$K_{\omega}=0.073$~\AA$^{-1}$),
then
$\delta_{\omega}$ so-deduced
remains non-zero up to the highest energy measured.
(Vertical thick bars.)
}
\label{fig:incomme}
\end{figure}
%=========================================================

Of particular interest is the energy scale of the dynamic
incommensurate correlations.  The scattering profiles evolve from a
well-resolved, double-peaked structure at low temperatures and
energies to a broad, single peaked structure with increasing $\omega$
and $T$.  At this stage, we cannot say with certainty that an
incommensurate-to-commensurate cross-over exists.  However, a simple
analysis of the data points in this direction.  If we assume that
$K_{\omega,T}$ evolves smoothly with $T$ as shown in the top panel of
Fig.~\ref{fig:incomme}, then 
the detailed fits revealed
that
$\delta_{\omega}$ goes to zero at around 
$\omega_{\,7\%}=20$~meV at
low $T$ as plotted in the lower panel. 
This energy scale of about 20 meV is reasonable in
light of similar incommensurate-to-commensurate cross-over behavior 
seen in other LSCO
compounds: $\omega_{\,2.4\%}=7$~meV~\cite{Matsuda-IC-C},
$\omega_{\,10\%}\approx 25$~meV~\cite{Petit}, and $\omega_{\,15\%
}>50$~meV.~\cite{Endoh-JPSJ} In addition, the temperature
dependence of $\chi^{\prime\prime}(\omega)_{1.5{\rm K}}$ changes slope
across $\omega$ of
20~meV as shown in Fig.~\ref{fig:S/chi_E}(b).  With increasing
temperature, $\delta_{\omega}$ decreases and appears to go to zero at
about 300~K ($\sim 25$~meV in energy units).  This consistency between
the crossover energy and temperature is also seen in
La$_{1.976}$Sr$_{0.024}$CuO$_{4}$.~\cite{Matsuda-IC-C} Alternatively,
the incommensurability $\delta_{\omega}$ may remain non-zero up to the
highest measured energy transfers, if we assume a constant
$\kappa_{\omega,T}=0.04~{\rm r.l.u.}$ for the whole energy range.
This restriction appears to be less likely, and the profiles shown in
Figs.~\ref{fig:q-spectra_E} and \ref{fig:q-spectra_T} are fit better if an
incommensurate-to-commensurate cross-over does indeed occur.

%\subsection{Speculation}
In superconducting La$_{1.93}$Sr$_{0.07}$CuO$_{4}$, the observations
of an enhancement of $\chi^{\prime\prime}(\omega)$ at low $\omega$,
the scaling function with a form of tan$^{-1}(\omega/{\rm k}_{\rm
  B}T)$, and the incommensurate-to-commensurate crossover at a finite
energy, are all reminiscent of the behavior observed in the
non-superconducting spin-glass
compositions.~\cite{Keimer,Matsuda-JPSJ,Matsuda-IC-C} Even though the
positions of the elastic incommensurate scattering changes
dramatically as the doping level crosses the
insulator-to-superconductor boundary,~\cite{Wakimoto-Direct}
it appears
that the behavior of the inelastic spin excitations evolves quite
smoothly.  This is consistent with the apparent 
observation of 
spin-glass ordering in the underdoped superconducting region observed
by $\mu$SR measurements.~\cite{msr-1,msr-2} Also, the saturation of
$\kappa_{\omega}^{\ 7\%}$ below 10~meV or 100~K 
in Fig.~\ref{fig:incomme}(a) is
similar to the behavior of the
inverse correlation lengths of spin-glass
compounds.~\cite{Keimer}

%========FIGURE INSERTION=================================
%\begin{figure}
%\centerline{\epsfxsize=3in\epsfbox{fig2.eps}}
%\caption{}
%\label{}
%\end{figure}
%=========================================================

\section{Conclusion}
We have conducted magnetic inelastic neutron scattering experiments on
La$_{1.93}$Sr$_{0.07}$CuO$_{4}$ 
over a wide range of $\omega$ and $T$.
The low-energy incommensurate fluctuations are enhanced at low $T$,
with no indication of a magnetic gap in this underdoped
superconductor.  The dependence of $\chi^{\prime\prime}(\omega)$ on
$T$ and $\omega$ 
is observed to depend primarily on the scaled valuable
($\omega/{\rm k}_{\rm B}T$).  By following the scattering profiles to
high energy, we conjecture that an incommensurate-to-commensurate
crossover occurs at $\omega \sim$ 20~meV or $T \sim 300$~K.  These
results are qualitatively similar to the behavior observed in the
non-superconducting spin-glass compositions, despite the fact that the
positions of the incommensurate peaks change drastically across the
insulator-superconductor boundary of the LSCO phase diagram.  The
behavior of this material La$_{1.93}$Sr$_{0.07}$CuO$_{4}$
is a crucial link in the chain of
understanding how the static and dynamic spin correlations evolve from
the insulator to the optimal superconductor.

\section{Acknowledgments}
We would like to thank G. Shirane, K. Yamada, M. Matsuda, S. Wakimoto,
K. Hirota, H. Kimura and M. A. Kastner
for fruitful discussions.  We also thank B.
Keimer, M.  Onodera, K. Nemoto, and L.P. Regnault for their assistance
in our experiments.  The present work was supported by the US-Japan
Cooperative Research Program on Neutron Scattering, a Grant-in-Aid for
Scientific Research of Monbusho, and the Core Research for Evolutional
Science and Technology (CREST) Project sponsored by the Japan Science
and Technology Corporation.
The work at MIT was supported by the U.S. National Science Foundation
under contract numbers DMR-0071256 and DMR98-08941.

%%%%%%%% main text (above) %%%%%%%%%%%%%%

%%%%%%%% references (below) %%%%%%%%%%%%%

%%%%%%%% references (above) %%%%%%%%%%%%%

\end{document}